# BİLGİ AĞLARI VE BİR BİLGİ AĞI ÖRNEĞİ

*Hilmi Bahadır Temur[1*], Ahmet Serdar Yılmaz[2], Mehmet Tekerek[3]*

[1]Kahramanmaraş Sütçüimam Üniversitesi, Enformatik Anabilim Dalı, 46040, Kahramanmaraş, Türkiye
[2]Kahramanmaraş Sütçüimam Üniversitesi, Elektrik-Elektronik Mühendisliği Bölümü 46040, Kahramanmaraş, Türkiye
[3]Kahramanmaraş Sütçüimam Üniversitesi, Eğitim Fakültesi, Bilgisayar ve Öğretim Teknolojileri Eğitimi Bölümü, 46040, Kahramanmaraş, Türkiye

*Sorumlu yazar:temurbahadir@gmail.com

**ÖZET**

*Bilgi ağları, insan zekâsının çalışması sonucu oluşan düşünsel ürün olarak tanımlanan bilginin, herhangi bir iletişim aracı ile başkasına aktarılmasını sağlayan sosyal ağlar olarak tanımlanabilir. Bir bilgi ağı, en üst düzeyde değer yaratmak için, öncelikle bilgi üretme ve aktarma süreci yoluyla bilgiyi biriktirmek ve kullanmak amacıyla bir araya getirilen çok sayıda insanı, kaynağı ve bunlar arasındaki ilişkileri temsil eder. Bilgi ağlarının genel yapısı; bilgiyi toplama, organize etme ve yayma olmak üzere üç temel aşamadan oluşur. İlk aşama olan bilgi toplama, kurum ve kuruluşlarda var olan bilginin ağ yapısına girme evresidir. Organize etme aşaması, ağ yapısına katılan düzensiz ve yapılandırılmamış bilgilerin belirli standartlara göre yapılandırılması ve düzenli bir şekilde yapıya kaydedilmesidir. Bilgi yayma ise organize edilmiş bilgilerin kullanıcı bilgi ve ihtiyaçları doğrultusunda aktarılması olarak ifade edilebilir. Bilgi ağlarının eğitim, ekonomi, sosyal ve kültürel alan gibi her alanda kullanımı ve etkileşimi artmaktadır. Bu çalışmada eğitim alanında bir bilgi ağı örneği oluşturulmuştur. Eğitim bilgi ağı örneğinde amaç, bir ders süreci içerisinde öğrenci, öğretmen ve kılavuz arasında iletişimi sağlamak ve oluşan bilgileri depolamak, sistemli bir şekilde kaydedilmesini sağlamak, kullanıcıların ihtiyaçları doğrultusunda yaymak ve bilgilerin güncellenmesi imkânını sağlamaktadır. Eğitim bilgi ağı için dinamik web sitesi ve içerik yönetim sistemi geliştirilmiştir. Web sitesinin yapımında HTML, JavaScript, PHP ve veri tabanı olarak MySQL tercih edilmiştir. Eğitim bilgi ağı, web teknolojilerini destekleyecek esnek bir yapıya sahiptir. Bilginin üretimini, yönetimini ve yayınlanmasını belirli ihtiyaçlar çerçevesinde gerçekleştirerek, farklı sorunlara, farklı ve duruma özel çözümler üreten bilgi ağı oluşumunu sağlamaktadır. Eğitim bilgi ağı örneği ile yenilikçi bilgi ağ yapısına uygun geliştirilebilir bir bilgi ağı tasarlanmıştır.*

**Anahtar Kelimeler:** Bilgi ağları, Bilgi paylaşımı, Bilgi yönetimi, Bir bilgi ağı örneği

# KNOWLEDGE NETWORK AND A KNOWLEDGE NETWORK EXAMPLE

**ABSTRACT**

*Knowledge networks can be defined as social networks that enable the transfer of the knowledge, which is defined as the intellectual product formed as a result of the work of human intelligence, to be transferred to any other means of communication. A knowledge network represents a large number of people, resources and relationships between them, to create the highest value, primarily to accumulate and use knowledge through the process of generating and transmitting knowledge. General structure of knowledge networks; it consists of three basic stages: gathering, organizing and disseminating knowledge. The first step, knowledge collection, institutions and organizations to enter the network structure of the knowledge that is present. The organizing phase is the structuring of irregular and unstructured knowledge in the network structure according to certain standards and recording them regularly in the structure. Knowledge dissemination can be expressed as the transfer of organized knowledge in accordance with user knowledge and needs. The use and interaction of knowledge networks in every field such as education, economy, social and cultural area is increasing. In this study, a knowledge network is created in the field of education. The purpose of the training knowledge network is to ensure communication between the student, the teacher and the guide, and to store the knowledge that is formed in a course, to enable the system to be recorded in a systematic way, to disseminate it according to the needs of the users and to update the knowledge. A dynamic website and content management system have been developed for the training knowledge network. In the construction of the website, HTML, JavaScript, PHP and MySQL as the database were preferred. The training knowledge network has a flexible structure to support web technologies. By making production, management and publication of knowledge within the framework of specific needs, it provides knowledge network for different problems, producing different and tailored solutions. With the example of education knowledge network, a knowledge network that can be developed according to the innovative knowledge network structure is designed.*

1. GİRİŞ

Bilgi ağı temel olarak; bilginin yönetimi, organizasyonu ve paylaşılmasını amaçlayan her türlü ağ olarak tanımlanır. Öte yandan, bilgi ağları gibi karmaşık ve çok yönlü bir kavramı tanımlamak genellikle zordur çünkü katılımcılar bu ağların faaliyetlerini gerçek ihtiyaçlara ve zorluklara cevap verecek şekilde sürekli olarak şekillendirir ve genişletir. Bu sebeple, bilgi ağının oluşturulması, geliştirilmesi ve kullanılması ile ilgili yeni yaklaşımları benimsemek ve desteklemek doğru bir adımdır (McKnight, Lee & Cukor, Peter).

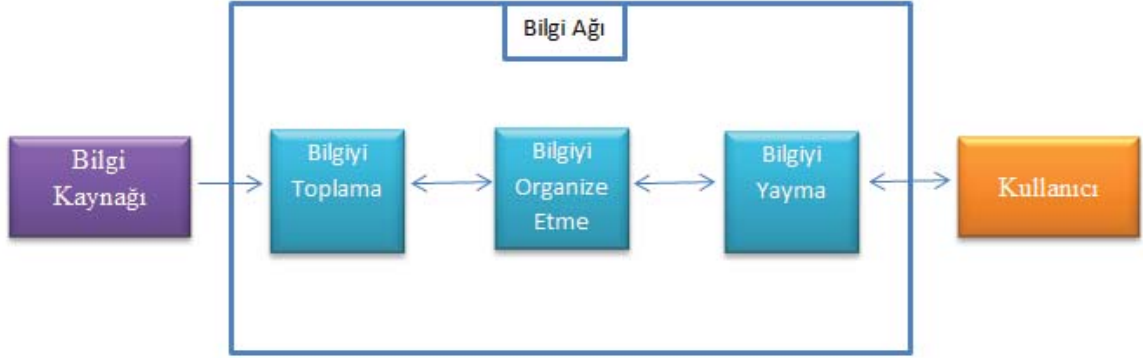

Şekil 1 Bilgi ağı genel yapısı

Önemli olan, yalnızca bir veri deposundaki gibi yığılma veri birikimi değil; bilgi üretiminin, paylaşımının ve iletiminin de teşvik edilmesidir(Phelps C, Heidl R, Wadhwa). Bilgi üretme ve aktarma; gerçek (örneğin ofiste, müşteri ile), sanal (örneğin, dağınık ekip odaları) veya zihinsel (örneğin ortak değerler, fikirler ve idealler) yerlerde gerçekleşebilir. Bilgi ağlarının temel işlevi, kurum içi veya kurumlar arası bilgileri edinerek paylaşmak ve erişilebilir kılmaktır. Ağ oluşturma, kuruluşların gerekli bilgileri bulmalarına ve başarılı yenilikleri gerçekleştirmek için kullanmalarına yardımcı olur. Bu nedenle, kurumlar bilgi ağlarını başlatarak bilgi paylaşımını artırmaya çalışırlar (Rezaeian, A., & Bagheri, R).

Allee, Bilgi ağlarının faydaları genel anlamda birey, topluluk ve işletmeler boyutunda ele alınarak incelenmiştir.

**İşletmeler İçin:**

• Strateji geliştirmeye yardımcı olur.

• Yerel ve kurum çapında daha hızlı problem çözmeyi destekler.

• Yetenek geliştirme, işe alma ve elde tutma konusunda yardımcı olur.

• Temel yetenekler ve bilgi yetkinlikleri oluşturur.

• Operasyonel mükemmellik için uygulamaları daha hızlı dağıtmaya olanak sağlar.

• İnovasyon için fırsatları arttırır.

**Topluluk için:**

• Belirli diller etrafında ortak dil, yöntem ve modeller oluşturulmasına yardımcı olur.

• Daha geniş bir nüfusa bilgi ve uzmanlık kazandırır.

• Çalışanlar şirketten ayrıldıklarında bilgilerin saklanmasına yardımcı olur.

• Şirket genelinde uzmanlığa erişimi arttırır.

• Gücü paylaşmak ve kuruluşun resmi bölümleriyle nüfuz etmenin bir yolunu sağlar.

**Birey için:**

• İnsanların işlerini yapmalarına yardımcı olur.

• Şirkette istikrarlı bir topluluk duygusu sağlar.

• Öğrenme odaklı bir kimlik duygusu geliştirir.

• Bireysel beceri ve yetkinlik geliştirmeye yardım eder.

• Bilgi çalışanının güncel kalmasına yardımcı olur.

Özellikle II. Dünya Savaşından sonra bilginin önemin farkına varılmasıyla bilgi ve bilgi ağları hakkında çalışmalar artmıştır. 1970 lerde internetin kullanımının yaygınlaşması bilgi ağlarının önemli derecede yayılmasına ve çok geniş kitlelere ulaşmasına olanak sağlamıştır. Araştırmacılar bilgi ve bilgi ağları hakkında hatırı sayılır derecede araştırma yapmaktadır. TÜBİTAK ULAKBİM çatısı altında, Türkiye'de yayımlanan akademik dergiler için elektronik ortamda barındırma ve editoryal süreç yönetimi hizmeti sunan dergipark.org.tr adresinden veriler incelendiğinde ülkemizde de bilgi ve bilgi ağları konularında artan sayıda çalışmalar yapılmaktadır.

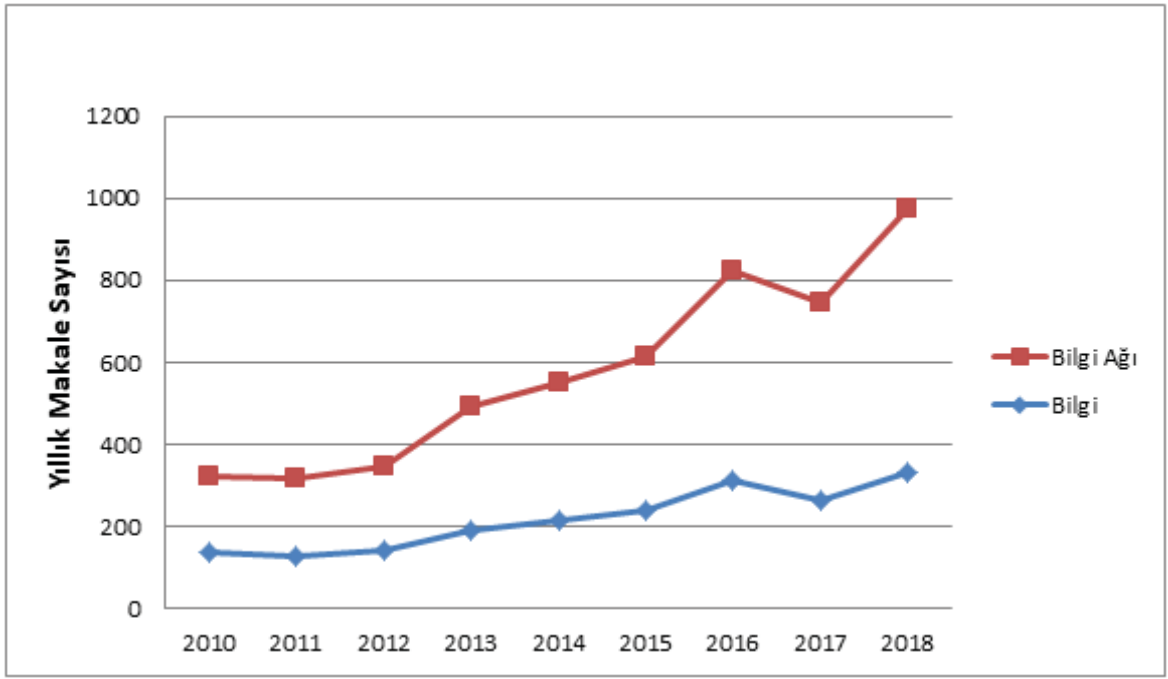

Şekil 2 Bilgi Ve Bilgi Ağları Hakkında Yapılan Çalışmaların Yıllara Göre Sayısı

Bu çalışmada bilgi ağlarının yapısının ve öneminin daha iyi anlaşılması için örnek bilgi ağı oluşturulmuştur.

## 2. MATERYAL VE YÖNTEM

Bu çalışmada bilgi ağlarının yapısının ve öneminin daha iyi anlaşılması için örnek bilgi ağı oluşturulmuştur. Oluşturulan Hücresel Bilgi Ağı için dinamik web sitesi tasarlanmış ve HTML, JavaScript, PHP ve veri tabanı olarak MySQL tercih edilmiştir.

### 2.1. Hücresel Bilgi Ağı Kullanıcı Türleri ve Özellikleri

Hücresel Bilgi Ağında sistemin yöneticisi admin, öğretmen, çaylak öğrenci ve kılavuzdan oluşmaktadır. Aşağıda kullanıcı görevleri yer almaktadır.
1. Admin:

   ✓ Sisteme yeni kaydolacak öğretmeni onaylar.
   ✓ Öğretmenin girdiği dersleri onaylar.

2. Öğretmen:

   ✓ Ders açar.
   ✓ Açılan derse çaylak öğrencilerin onayını yapar.

- ✓ Kılavuz ile çaylağın eşleştirmesini onaylar.
- ✓ Ödev verdiği öğrencilerin sorduğu sorulara cevap verir.
- ✓ Kılavuz ve çaylak arasındaki yazışmaları görüp, ödevde hangi aşamada olduklarını ve hatalarını anlık görebilir.
- ✓ Başarılı olan öğrencilerin ödevini değerlendirir.

3. Kılavuz:

İkiye ayrılır:
a) Öğrenci: Eğitim hayatına devam eden ve daha önceden dersi başarı ile geçmiş gruptur.
b) Mezun: Okuldan mezun olmuş ve iş hayatında olan gruptur, bu grup ayrıca iş hayatının içinde aktif olan, tecrübe sahibi kişilerdir.

- ✓ Ödevlerde öğrencilere rehberlik eder.
- ✓ Öğrenci ile ödev arasında bir aracı görevi görür.
- ✓ Çaylaklardan gelen kılavuz olma isteğine cevap verir, istediği kişiye kılavuz olabilir.
- ✓ Çaylak tarafından gönderilen ödevleri değerlendirir.

4. Çaylak:

- ✓ Derse çaylak olarak kaydolur.
- ✓ Kılavuzunu seçer .
- ✓ Kılavuz dönütlerini ödevine uygular ve öğretmene son halini gönderir.

## 2.2. Hücresel Bilgi Ağı Senaryosu

1. Adım: Kişiler sisteme kayıt olacak, eğer kişi;
    a)Öğretmen ise, akademik olarak verdiği dersler, uzmanlık alanları istenecek.
    b)Öğrenci ise, önceki almış ve geçmiş olduğu dersler ve notu istenecek.
    c)Mezun ise, sadece önceden almış olduğu dersler ve harf notu istenecek.
2. Adım: Öğretmen ders açmak için talepte bulunacak.
3. Adım: Admin ders onaylayacak.
4. Adım: Öğrenciler ders seçecek ve 5 kılavuz adayına kılavuzluk daveti gönderilecek.
5. Adım: Davet alan kılavuzlar bir ders için en fazla 5 çaylak davetini kabul edebilecek.
6. Adım: Çaylağın davetini kabul eden kılavuzlardan birini seçecek.
7. Adım: Tüm Çaylak- Kılavuz eşleşmesi sağlandıktan sonra öğretmen ders modülüne ders materyallerini paylaşacak.
8. Adım: Öğretmen tarafından ödev verilecek.
9. Adım: Dersi alan öğrenci kılavuzu rehberliğinde ödevi yapacak.
10. Adım: Ödev tamamlandıktan sonra Kılavuz ödev hakkında görüşünü ve değerlendirmesini yapacak.
11. Adım: Kılavuz değerlendirmesinde sonra öğretmene iletilecek.
12. Adım: Öğretmen ödevi değerlendirecek.
13. Adım: Ders modülü bitene kadar öğretmen tarafından tekrar ödev verilerek 8. Adıma geçilecek.
14. Adım: Ders modülü tamamlandıktan sonra öğretmen ders modülünü kapatacak.
15. Adım: Öğrenci istemesi halinde Portfolyo butonundan öğrencinin yapmış olduğu tüm çalışmaların dökümüne erişebilecek.
16. Adım: Her yeni dönemde kişi bilgileri güncellenerek sistem 1. Adım a tekrar geçebilecek.

## 2.3. Hücresel Bilgi Ağı Ekran Görüntüleri

Şekil 3 Öğretmen başvuru formu

Şekil 4 Ders modülü

Ders modülündeki Ana sayfa menüsünde dersin adı, içeriği, dersin başlama ve bitiş tarihleri yer almaktadır. Kılavuz menüsünde çaylak öğrencilerin için derste rehberlik etmeleri için daha önce bu dersten başarılı öğrenciler arasında kendilerine kılavuz seçmeleri sağlanmaktadır. Duyurular menüsü öğretmenin haberleşme alanıdır. Ödevler menüsünde öğretmenin ödev göndermesi, gelen ödevlerin değerlendirmesini yapmaktadır. Öğrenciler ise gelen ödevleri teslim ettiği ve kılavuzu ile ödev hakkında mesajlaştığı alanlar mevcuttur. Tartışma menüsünde öğrenciler için istedikleri konular hakkında tartışma alanı sağlamaktadır. Notlar kısmında herkesin ödev değerlendirme sonuçları ve ortalamaları mevcuttur. Katılımcılar menüsünde kayıtlı öğrenciler ve kılavuzları görüntülenmektedir. Ders programı menüsüne öğretmenin ders materyali paylaşması sağlanmaktadır. Son olarak çaylak öğrencilerin öğretmen ile iletişimi sağlamak amacıyla öğretmene sor menüsü bulunmaktadır. Web sitesinin ana menüsünde bulunan portfolyo alanıyla öğrenciler kendi çalışmaları ve diğer kullanıcıların çalışmalarına erişebilmektedirler.

## 3. SONUÇ

Bu çalışmada, bilgi, ağ ve bilgi ağı kavramları ile bilgi ağlarının yapısından, faydalarından ve öneminden bahsedilmiştir ve örnek bilgi ağı oluşturulmuştur. Oluşturulan Hücresel Bilgi Ağı ile bir ders sürecinde meydana gelen bilgilerin toplanması, toplanan bilgilerin sistemli bir şekilde depolanması ve kullanıcı ihtiyaçları doğrultusunda bilgilerin iletilmesi aşamaları aktarılmıştır.

Hücresel Bilgi Ağının en önemli özelliği, eğitim öğretim sürecinde öğrencilere rehberlik edecek, daha önce alınan dersten başarılı olan kılavuz desteği sağlamasıdır. Bu sayede, öğrencilerin kendilerini geliştirmesi hedeflenmektedir. Yine kılavuz sayesinde ödevlerin daha özgün ve etkili olması sağlanmaktadır. Ödevlerin öğretmenden önce kılavuz tarafından değerlendirmesi de öğretmenin ödevi değerlendirmesine yardımcı olmaktadır. Ağda mevcut olan, daha önceden yapılmış çalışmalar sayesinde öğrencilerin literatür taraması ihtiyacını da karşılamaktadır.

Hücresel Bilgi ağı, web ortamında kullanılabilecek farklı tipte dokümanların ve dosyaların yönetimine olanak sağlamaktadır. Kullanıcının kendi ilerlemesini görebileceği bir portfolyo hizmeti sunmaktadır. Gelişen web teknolojilerini destekleyecek esnek ve geliştirilebilir bir alt yapıya sahiptir. Bilginin üretimini, yönetimini ve yayınlanmasını belirli ihtiyaçlar çerçevesinde gerçekleştirerek, farklı sorunlara, farklı ve duruma özel çözümler sağlamaktadır.

## 4. KAYNAKLAR